\documentclass[aps,twocolumn]{revtex4}
\usepackage{graphicx,epsfig}
\usepackage[usenames]{color}
\usepackage{amsmath}
\date{\today}

\newcommand{\bi}{{\bf i}}
\newcommand{\bj}{{\bf j}}

\newcommand{\cV}{{\mathcal V}}
\newcommand{\cZ}{{\mathcal Z}}

\begin{document}
%
%\title{Correction to the Fermi liquid theory by fractional exclusion statistics}
\title{The Fermi liquid theory with fractional exclusion statistics}
\author{Drago\c s-Victor Anghel}
\affiliation{Horia Hulubei National Institute for Physics and Nuclear Engineering, 30 Reactorului street, P.O.BOX MG-6, M\u agurele, Jud. Ilfov, Romania}
%
%\pacs{05.30.-d}{Quantum statistical mechanics}
%\pacs{05.30.Ch}{Quantum ensemble theory}
%\pacs{05.30.Pr}{Fractional statistics systems (anyons, etc.)}
%
% \pacs{05.30.-d}{Quantum statistical mechanics}
% \pacs{05.30.Ch}{Quantum ensemble theory}
% \pacs{05.30.Pr}{Fractional statistics systems (anyons, etc.)}
%
\begin{abstract}

The Fermi liquid theory may provide a good description of the thermodynamic properties of an interacting particle system when the interaction between the particles contributes to the total energy of the system with a quantity which may depend on the total particle number, but does not depend on the temperature. In such a situation, the ideal part of the Hamiltonian, i.e. the energy of the system without the interaction energy, also provides a good description of the system's thermodynamics. 

If the total interaction energy of the system, being a complicated function of the particle populations, is temperature dependent, then the Landau's quasiparticle gas cannot describe accurately the thermodynamics of the system.

A general solution to this problem is presented in this paper, in which the quasiparticle energies are redefined in such a way that the total energy of the system is identical to the sum of the energies of the quasiparticles. This implies also that the thermodynamic properties of the system and those of the quasiparticle gas are identical. 

By choosing a perspective in which the quasiparticle energies are fixed while the density of states along the quasiparticle axis vary, we transform our quasiparticle system into an ideal gas which obey fractional exclusion statistics. 
\end{abstract}

\maketitle

\section{Introduction}\label{intro}

%Landau's Fermi liquid theory (FLT) \cite{PinesNozieres:book} describes systems of interacting bosons or fermions by making use of a gas of properly defined quasiparticles. The expresion of the population of the quasiparticle states is the same as the expression of the ideal Bose or Fermi populations by the ideal (Bose or Fermi) populations

In the Fermi liquid theory (FLT) \cite{PinesNozieres:book} a system of interacting bosons or fermions is described as a gas of quasiparticles. The population of the quasiparticle states has the expression of the ideal Bose or Fermi population (depending on the type of particles that are in the system) if the quasiparticle energy, $\tilde\epsilon^L$, is defined as the variation of the total energy of the system with respect to the population of the quasiparticle state. %The occupation of the quasiparticle states influence in general the quasiparticle energies. 

If the system is not trivial, the sum of the energies of the quasiparticles is not equal to the total energy of the system and since the quasiparticle energies also depend on the occupation of the quasiparticle states, the thermodynamic properties of the quasiparticle gas may neither be similar to those of an ideal particle gas, nor to the thermodynamic properties of the original system. In such a case, the ideal gas characteristics of the interacting particle system, conjectured based on an idealized quasiparticle model (e.g. the textbook result that $C_V\propto T$ in the low temperature limit \cite{PinesNozieres:book}) is not rigorously justified.

In this paper I propose a method to describe the FLT system as an ideal gas of fractional exclusion statistics (FES) particles \cite{PhysRevLett.67.937.1991.Haldane}. For this I redefine the quasiparticle energies such that the total energy of the system equals the energy of the quasiparticle gas, ensuring in this way that the thermodynamic properties of the system are identical to the thermodynamic properties of the quasiparticle gas. Moreover, %while Landau's quasiparticle energies, $\tilde\epsilon^L$, depend on the population of the other quasiparticle states, 
in this method the quasiparticle energies do not depend on the occupations of the quasiparticle states and therefore \textit{the gas is ideal}, unlike Landau's quasiparticle gas. 

The FES has been applied before to FLT type of systems, but with equally spaced free-particle energy levels, $\epsilon_\bi$, and the particle-particle interaction described in the mean field approximation -- i.e. $\sum_{\bi\bj}V_{\bi\bj}n_\bi n_\bj/2\equiv VN(N-1)/2$ \cite{PhysRevLett.73.3331.1994.Murthy,NuclPhysB470.291.1996.Hansson,IntJModPhysA12.1895.1997.Isakov,JPhysB33.3895.2000.Bhaduri,PhysRevLett.86.2930.2001.Hansson,RJP.54.281.2009.Anghel}. Other types of closely related systems to which FES has been applied are the Bose and Fermi gases described in the thermodynamic Bethe ansatz  \cite{PhysRevLett.73.2150.1994.Isakov,NewDevIntSys.1995.Bernard,PhysRevB.56.4422.1997.Sutherland,PhysRevLett.85.2781.2000.Iguchi,PhysRevE.75.61120.2007.Potter,PhysRevE.76.61112.2007.Potter,EPL.90.10006.2010.Anghel} and spin chains \cite{PhysRevE.84.021136.2011.Liu,PhysRevE.85.011144.2012.Liu}

The systems of equal and constant DOS (i.e. equally spaced free-particle energy levels) described in the mean-field approximation are all thermodynamically equivalent (i.e. they all have the same heat capacity and entropy for any temperature and particle number), no matter if the particles are bosons, fermions, or obey FES of constant, diagonal FES parameters \cite{JPA35.7255.2002.Anghel}. Therefore the FES is not a necessary tool for such systems. 

The FES has been applied to systems of any DOS and general type of interaction in Ref. \cite{PhysLettA.372.5745.2008.Anghel,PhysLettA.376.892.2012.Anghel} and this provides the method for describing systems of interacting particles as ideal gases that I shall use in this paper.

The paper is organized as follows. In Section \ref{basics} I present briefly the basics of FLT, to specify the notations. In Section \ref{section_discrepFLT} I show the limitations of the Landau's quasiparticle model in the description of the interacting particle system. In Section \ref{section_FES} I introduce the description of the system based on the ideal FES gas. 

\section{Basics quantities of FLT}\label{basics}

In the FLT the system is described by the general energy functional \cite{PinesNozieres:book},
\begin{equation}
E(\{n_\bi\}) = \sum_\bi\epsilon_\bi n_\bi + \frac{1}{2}\sum_{\bi\bj}V_{\bi\bj}n_\bi n_\bj, \label{Etotgen}
\end{equation}
where the subscripts $\bi$ and $\bj$ represent single-particle quantum numbers, $n_\bi$, $n_\bj$, are occupation numbers, whereas $V_{\bi\bj}$ is a phenomenological term which describes the particle-particle interaction. 

For concreteness I discuss only fermions. Then the grandcanonical partition function is
\begin{eqnarray}
 \log(\cZ) &=& -\sum_\bi[n_\bi\log(n_\bi) + (1-n_\bi)\log(1-n_\bi)] \nonumber \\ 
  && - \beta[E(\{n_\bi\})-\mu N] \label{cZ_fermi}
\end{eqnarray}
and the (average) populations, denoted by $f_\bi$, are obtained by imposing the maximization condition, $\partial\cZ(\{f_\bj\})/\partial f_\bi=0$. This gives
\begin{equation}
 f_\bi(T,\mu) \equiv f(T,\mu,\tilde\epsilon_\bi^L) = \left[e^{\beta(\tilde\epsilon_\bi^L-\mu)}+1\right]^{-1}, \label{ni_fermi}
\end{equation}
where 
\begin{equation}
 \tilde\epsilon_\bi^L(T,\mu) \equiv \left.\frac{\partial E(\{n_\bj\})}{\partial n_\bi}\right|_{n_\bj=f_bj} = \epsilon_\bi + \sum_jV_{\bi\bj}f_\bj(T,\mu) \label{epsilonL}
\end{equation}
is the Landau's quasiparticle energy. I use the typical notation, $\beta=1/(k_BT)$, where $T$ is the temperature and $k_B$ is the Boltzmann's constant.

At equilibrium, the internal energy, the total particle number and the entropy of the system are
\begin{equation}
  U(T,\mu) = \sum_\bi\epsilon_\bi(T,\mu) f_\bi(T,\mu) + \frac{1}{2}\sum_{\bi\bj}V_{\bi\bj}f_\bi(T,\mu) f_\bj(T,\mu), \label{U_gen}
\end{equation}
\begin{equation}
 N(T,\mu) = \sum_\bi f_\bi(T,\mu), \label{rel_mu}
\end{equation}
and
\begin{eqnarray}
  S(T,\mu) &=& -k_B\sum_\bi\{[1-f_\bi(T,\mu)]\log[1- f_\bi(T,\mu)] \nonumber \\ 
  && + f_\bi(T,\mu)\log[f_\bi(T,\mu)]\} \label{S_gen}
\end{eqnarray}
respectively. 

%We shall see in this paper that the thermodynamics of the interacting fermions system described by Eq. (\ref{Etotgen}) cannot be formulated as the thermodynamics of an ideal Fermi gas except in simple cases and, rigorously speaking, the linear temperature dependence of the heat capacity of the system in the low temperature limit, does not follow from the thermodynamics of the quasiparticle gas in general conditions even in the so called \textit{range of applicability of the FLT}. To describe a gas of interacting fermions as an ideal gas, one should use ideal particles which obey \textit{fractional exclusion statistics} (FES).

\section{Thermodynamics in the FLT}\label{section_discrepFLT}

\subsection{Internal energy difference}\label{subs_U_diff}

In what follows I shall call the \textit{Landau's gas} (LG) a gas of quasiparticles of energies $\tilde\epsilon_\bi^L$ (\ref{epsilonL}) and of total energy
\begin{equation}
  E^{L}(\{n_\bi\}) \equiv \sum_\bi\tilde\epsilon_\bi^L n_\bi = E(\{n_\bi\}) + \frac{1}{2}\sum_{\bi\bj}V_{\bi\bj}n_\bi n_\bj . \label{sum_teps}
\end{equation}
The equilibrium $E^{L}$ is denoted by $U^L(T,\mu)= U(T,\mu) +\sum_{\bi\bj}V_{\bi\bj}f_\bi f_\bj/2$. We observe the difference, $\cV(T,\mu) =U^L(T,\mu) -U(T,\mu)= \sum_{\bi\bj}V_{\bi\bj}f_\bi(T,\mu)f_\bj(T,\mu)/2$. 

I shall say that one can use the LG model to calculate the thermodynamic properties of the original system if the two of them are thermodynamically equivalent, i.e. $C_V(T,N)\equiv\partial U(T,N)/\partial T = C_V^L(T,N) \equiv\partial U^L(T,N)/\partial T$. This implies that $\cV(T,N)\equiv\cV(N)$ is just a function of $N$. 

Furthermore, if $\cV(T,N)\equiv\cV(N)$, then the gas of noninteracting particles, of total energy 
\begin{equation}
  U^\text{id}(T,\mu) = \sum_\bi\epsilon_\bi f_\bi= U(T,\mu)-\cV(N)= U^L(T,\mu)-2\cV(N), \label{Etot_id}
\end{equation}
is also thermodynamically equivalent with both, the original gas and the LG:
\begin{equation}
 C_V^\text{id}(T,N)=C_V^L(T,N)=C_V(T,N), \label{equal3CV}
\end{equation}
where $C_V^\text{id}(T,N)\equiv\partial U^\text{id}(T,N)/\partial T$. This is a typical case of system described in the mean field approximation, in which the interaction energy -- dependent on $N$ -- contributes only as a background energy. 

If all the three systems are equivalent -- Eqs. (\ref{equal3CV}) is satisfied -- then the application of the FLT is trivial and not necessary; one should better work with the ideal gas, to which the background interaction energy should be added. 

Reversely, if we assume by \textit{reductio ad absurdum} that $C_V(T,N)$ is different than $C_V^\text{id}(T,N)$, but it is well approximated by $C_V^L(T,N)$, we obtain a contradiction with Eq. (\ref{equal3CV}). Therefore either the thermodynamics is determined by the ideal particle description and the application of the FLT is trivial or it gives a wrong result and therefore there $U^L(T,\mu)$ should not be used for thermodynamics calculations. 

\subsection{The heat capacity of the FLT gas}\label{subs_CV_FES}

In general, the heat capacity of a quasiparticle gas, like the LG, is not that of an ideal Fermi gas due to the dependence of the quasiparticle energies on the populations of the single-particle states. 

The heat capacity of any system may be calculated by either of the formulas
\begin{subequations} \label{CVs_gen}
\begin{equation}
  C_V \equiv \left(\frac{\partial U}{\partial T}\right)_N = \left(\frac{\partial U}{\partial T}\right)_\mu - \left(\frac{\partial U}{\partial\mu}\right)_T\left(\frac{\partial N}{\partial T}\right)_\mu\left(\frac{\partial N}{\partial\mu}\right)_T^{-1}. \label{CV_gen_U}
\end{equation}
or 
\begin{equation}
  \frac{C_V}{T} \equiv \left(\frac{\partial S}{\partial T}\right)_N = \left(\frac{\partial S}{\partial T}\right)_\mu - \left(\frac{\partial S}{\partial\mu}\right)_T\left(\frac{\partial N}{\partial T}\right)_\mu\left(\frac{\partial N}{\partial\mu}\right)_T^{-1}. \label{CV_gen_S}
\end{equation}
\end{subequations}
where $S$ is the entropy of the system. 

For an ideal Fermi gas, $U^{id}(T,\mu)=\sum_\bi\epsilon_\bi f_\bi$ and $S=-k_B\sum_\bi[f_\bi\log(f_\bi) + (1-f_\bi)\log(1-f_\bi)]$, so
\begin{subequations}\label{dUN_dTmu}
\begin{eqnarray}
  \frac{\partial U^{id}}{\partial T} &=& \sum_\bi\epsilon_i\frac{\partial f_\bi}{\partial T},\qquad 
  \frac{\partial U^{id}}{\partial\mu} = \sum_\bi\epsilon_i\frac{\partial f_\bi}{\partial\mu}, \label{dU_dTmu}
\end{eqnarray}
\begin{eqnarray}
  \frac{\partial S}{\partial T} &=& k_B\sum_\bi\ln\left(\frac{1-f_\bi}{f_\bi}\right) \frac{\partial f_\bi}{\partial T}, \nonumber \\
  \frac{\partial S}{\partial\mu} &=& k_B\sum_\bi\ln\left(\frac{1-f_\bi}{f_\bi}\right) \frac{\partial f_\bi}{\partial\mu},\label{dS_dTmu}
\end{eqnarray}
\begin{eqnarray}
  \frac{\partial N}{\partial T} &=& \sum_\bi\frac{\partial f_\bi}{\partial T},\qquad
  \frac{\partial N}{\partial\mu} = \sum_\bi\frac{\partial f_\bi}{\partial\mu}. \label{dN_dmu}
\end{eqnarray}
\end{subequations}
In the continuous limit I introduce the density of states (DOS), $\sigma(\epsilon)$, and, if we assume that $\sigma(\epsilon)\equiv C\epsilon^s$, where $C$ and $s$ ($s>-1$) are constants, then 
\begin{subequations}\label{UNCV_ideal}
\begin{eqnarray}
  U^{id} &=& C(k_BT)^{s+2}\Gamma(s+2)Li_{s+2}(-e^{\beta\mu}),\label{U_ideal} \\
  N &=& C(k_BT)^{s+1}\Gamma(s+1)Li_{s+1}(-e^{\beta\mu}),\label{N_ideal} \\
  \frac{C_V^{id}}{k_B} &=& C(k_BT)^{s+1}\Gamma(s+2)\left[(s+2)Li_{s+2}(-e^{\beta\mu}) \right. \nonumber \\
  && \left. - (s+1)\frac{Li_{s+1}(-e^{\beta\mu})}{Li_{s}(-e^{\beta\mu})}\right]; \label{CV_ideal}
\end{eqnarray}
\end{subequations}
$Li_n(x)$ is the polylogarithmic function \cite{Lewin:book,ActaPhysicaPolonicaB.40.1279.2009.Lee}. 

Using the asymptotic expansions,
\begin{subequations} \label{Li_n10_limT0}
\begin{eqnarray}
  Li_n(-e^{\beta\mu})&\stackrel{\beta\mu\gg1}{\approx}&\frac{(\beta\mu)^n}{\Gamma(n+1)}\left[1+\frac{\pi^2}{6}\frac{n(n-1)}{(\beta\mu)^2}\right], \nonumber \\
  && \text{for }n>1, \label{Li_n_limT0} \\
  Li_1(-e^{\beta\mu})&\stackrel{\beta\mu\gg1}{\approx}& \beta\mu+e^{-\beta\mu} , \label{Li_1_limT0} \\
  Li_0(-e^{\beta\mu})&\stackrel{\beta\mu\gg1}{\approx}& 1 - e^{-\beta\mu} , \label{Li_0_limT0} 
\end{eqnarray}
\end{subequations}
one obtains the standard low $T$ result,
\begin{equation}
  \frac{C^{(id)}_V}{k_B} = \frac{\pi^2}{3}k_BT\sigma(\mu), \label{CV_univ_L}
\end{equation}
where $\mu=\epsilon_F$ is the Fermi energy.

The same results, (\ref{CV_ideal}) and (\ref{CV_univ_L}), are obtained if one starts from Eq.  (\ref{CV_gen_S}). 

Equation (\ref{CV_univ_L}) is valid for any DOS, as long as $\sigma(\epsilon)$ is finite and continuous at $\epsilon_F$ \cite{Stone:book}. 

We should also note here that in general $C^{(id)}_V\ne(\partial U^{id}/\partial T)_\mu$ even in the low temperature limit. As we can immediately check from Eqs. (\ref{CV_ideal}) and (\ref{Li_n10_limT0}), $C^{(id)}_V=(\partial U^{id}/\partial T)_\mu$ is valid only when $s=0$, or, for a more general DOS, when $\sigma(\epsilon)$ is constant in a finite interval around $\epsilon_F$, but which is large enough as compared to $k_BT$.

I showed in the previous section that $C_V^L\ne C_V$, except in trivial cases, therefore Eq. (\ref{CV_gen_U}), with $U$ replaced by $U^L$, cannot be used to calculate $C_V$. Moreover, even if one would (wrongly) conjecture that $C_V^L=C_V$, Eq. (\ref{CV_ideal}) would not be valid because $\tilde\epsilon^L$ are not constant -- they depend on $T$ and $\mu$ through $f_i(T,\mu)$ (\ref{epsilonL}) -- and we have
\begin{eqnarray}
  \frac{\partial U^L}{\partial T} &=& \sum_\bi\left[\frac{\partial\tilde\epsilon_i^L}{\partial T}f_\bi+\tilde\epsilon_i^L\frac{\partial f_\bi}{\partial T}\right], \nonumber \\
  \frac{\partial U^{id}}{\partial\mu} &=& \sum_\bi\left[\frac{\partial\tilde\epsilon_i^L}{\partial\mu}f_\bi + \tilde\epsilon_i^L\frac{\partial f_\bi}{\partial\mu}\right]. \label{dUL_dTmu}
\end{eqnarray}
Therefore, based on $U^L$ one cannot infer anything rigorously, on general grounds, about the low temperature expression of $C_V$. 

Another possibility is to employ Eq. (\ref{CV_gen_S}) \cite{PRA7.304.1973.Pethic}, since the expression of the entropy of the quasiparticle gas, say $S^L$, is identical to the one of an ideal gas, namely 
\begin{equation}
  S^L(T,\mu) = -k_B\sum_\bi[f_\bi\log(f_\bi) + (1-f_\bi)\log(1-f_\bi)] \label{SL_gen}
\end{equation}

But this method does not lead to Eq. (\ref{CV_univ_L}) either, as we shall see now. From Eq. (\ref{epsilonL}) I get 
\begin{subequations}\label{dteps_dTdmu}
\begin{eqnarray}
  \frac{\partial\tilde\epsilon_\bi^L(T,\mu)}{\partial(k_BT)} %&=& \sum_\bj\frac{\partial\tilde\epsilon_\bi^L(T,\mu)}{\partial n_\bj} \frac{\partial n_\bj^L(T,\mu)}{\partial(k_BT)} \nonumber \\ 
  &=& \sum_\bj V_{\bi\bj} \frac{\partial f_\bj(T,\mu)}{\partial(k_BT)}, \label{dteps_dT} \\ 
  \frac{\partial\tilde\epsilon_\bi^L(T,\mu)}{\partial\mu} %&=& \sum_\bj\frac{\partial\tilde\epsilon_\bi^L(T,\mu)}{\partial n_\bj} \frac{\partial n_\bj^L(T,\mu)}{\partial\mu}  \nonumber \\ 
  &=& \sum_\bj V_{\bi\bj} \frac{\partial f_\bj(T,\mu)}{\partial\mu} , \label{dteps_dmu}
\end{eqnarray}
\end{subequations}
and therefore
\begin{subequations} \label{dni_dTdmu}
\begin{equation}
  \frac{\partial f_\bi^L(T,\mu)}{\partial(k_BT)} = \frac{\partial f_\bi(T,\mu,\tilde\epsilon_\bi^L)}{\partial(k_BT)} + \frac{\partial f_\bi(T,\mu,\tilde\epsilon_\bi^L)}{\partial\tilde\epsilon_\bi^L} \frac{\partial\tilde\epsilon_\bi^L(T,\mu)}{\partial(k_BT)} , \label{dni_dT}
\end{equation}
\begin{equation}
  \frac{\partial f_\bi^L(T,\mu)}{\partial\mu} = \frac{\partial f_\bi(T,\mu,\tilde\epsilon_\bi^L)}{\partial\mu} + \frac{\partial f_\bi(T,\mu,\tilde\epsilon_\bi^L)}{\partial\tilde\epsilon_\bi^L} \frac{\partial\tilde\epsilon_\bi^L(T,\mu)}{\partial\mu} . \label{dni_dmu}
\end{equation}
\end{subequations}
In Eqs. (\ref{dni_dTdmu}), 
\begin{subequations}\label{dn_idTdmu_id}
\begin{equation}
  \frac{\partial f(T,\mu,\tilde\epsilon_\bi^L)}{\partial(k_BT)} = \frac{\beta^2(\tilde\epsilon_\bi^L-\mu) e^{\beta(\tilde\epsilon_\bi^L-\mu)}}{\left[e^{\beta(\tilde\epsilon_\bi^L-\mu)}+1\right]^2} , \label{dn_idT_id}
\end{equation}
%
%and
%
\begin{equation}
  \frac{\partial f(T,\mu,\tilde\epsilon_\bi^L)}{\partial\mu} = \frac{\beta e^{\beta(\tilde\epsilon_\bi^L-\mu)}}{\left[e^{\beta(\tilde\epsilon_\bi^L-\mu)}+1\right]^2} \label{dn_idmu_id}
\end{equation}
\end{subequations}
are the ideal gas expressions.

Combining Eqs. (\ref{dteps_dTdmu}) and (\ref{dni_dTdmu}) I obtain the systems of equations for $\partial f_\bi(T,\mu)/\partial(k_BT)$ and $\partial f_\bi(T,\mu)/\partial\mu$,
\begin{subequations}\label{dn_idTdmu}
\begin{eqnarray}
  \frac{\partial f_\bi(T,\mu)}{\partial(k_BT)} &=& \frac{\beta^2(\tilde\epsilon_i^L-\mu) e^{\beta(\tilde\epsilon_i^L-\mu)}}{\left[e^{\beta(\tilde\epsilon_i^L-\mu)}+1\right]^2}\left[ 1 - \frac{1}{\beta(\tilde\epsilon_i^L-\mu)}\right. \nonumber \\ 
  && \left.\times\sum_j V_{\bi\bj}\frac{\partial f_\bj(T,\mu)}{\partial(k_BT)}\right], \label{dn_idT}
\end{eqnarray}
\begin{equation}
  \frac{\partial f_\bi(T,\mu)}{\partial\mu} = \frac{\beta e^{\beta(\tilde\epsilon_i^L-\mu)}}{\left[e^{\beta(\tilde\epsilon_i^L-\mu)}+1\right]^2}\left[ 1 - \sum_j V_{\bi\bj}\frac{\partial f_\bj(T,\mu)}{\partial\mu}\right]. \label{dn_idmu}
\end{equation}
\end{subequations}
If the summations in Eqs. (\ref{dn_idTdmu}) are equal to zero, we re-obtain the ideal gas results (\ref{dn_idTdmu_id}), which, if plugged into Eqs. (\ref{dS_dTmu}) and (\ref{dN_dmu}) lead, through Eq. (\ref{CV_gen_S}), to the low temperature expression (\ref{CV_ideal}). But this seems to be a very special case and in general the quasiparticle gas cannot be employed to calculate the contribution of the interaction terms to the thermodynamic properties of the system, especially to the calculation of the low temperature expression of $C_V$.

\section{The ideal FES gas}\label{section_FES}

%\subsection{The ideal gas}

The method by which one can describe as an ideal gas the general interacting particle system of energy given by Eq. (\ref{Etotgen}) is provided by the FES. 

%The purpose of Ref. \cite{PhysLettA.372.5745.2008.Anghel} was to define a system of type (\ref{Etotgen}) as a system of quasiparticles, in such a way that the total energy of the system is the sum of the energies of the quasiparticles. If we can define such quasiparticles, then the quasiparticle gas is thermodynamically equivalent to the original system at any temperature. 

For this I redefine the quasiparticle energies in such a way that the gas of quasiparticles is thermodynamically equivalent with the interacting particle gas at \textit{any temperature}, the energies of the quasiparticles do not depend on the occupation of the other quasiparticle levels and, as a consequence, the population of one quasiparticle level does not depend on the populations of the other quasiparticle levels. 

For the simplicity of the presentation, let's assume that the single particle quantum numbers, $\bi$, are the single-particle energies, $\epsilon_i$, of the free fermions and $\epsilon_i\le\epsilon_j$ if $i<j$. The quasiparticle energies are \cite{PhysLettA.372.5745.2008.Anghel,PhysLettA.376.892.2012.Anghel}
\begin{equation}
  \tilde\epsilon_i = \epsilon_i + \sum_{j=0}^{i-1} V_{ij}n_j , \label{epstilgen}
\end{equation}
which satisfy identically 
\begin{equation}
  E(\{n_i\}) = \sum_i n_i\tilde\epsilon_i . \label{cond_entot}
\end{equation}

In the continuous limit I introduce the free particles DOS, $\sigma(\epsilon)$, and the quasiparticle DOS, $\tilde\sigma(\tilde\epsilon)$. Like in Ref. \cite{PhysLettA.372.5745.2008.Anghel,PhysLettA.376.892.2012.Anghel}, I assume that the function $\tilde\epsilon(\epsilon)$ is bijective, so I can freely inverse it -- $\epsilon(\tilde\epsilon)$. Equation (\ref{epstilgen}) becomes
\begin{eqnarray}
\tilde\epsilon &=& \epsilon + \int_0^\epsilon V(\epsilon,\epsilon') 
\sigma(\epsilon')n(\epsilon')\,d\epsilon' \equiv \epsilon(\tilde\epsilon) \nonumber \\ 
  && + \int_0^{\tilde\epsilon} V[\epsilon(\tilde\epsilon),\epsilon(\tilde{\epsilon'})] \tilde\sigma(\tilde{\epsilon'})n(\tilde{\epsilon'})\,d\tilde{\epsilon'}. \label{epstilgenint}
\end{eqnarray}
%

%Let's see now where is the ideal gas, how FES is manifesting in this system and what is the advantage of using this formalism. 

\begin{figure}
  \centering
  \includegraphics[width=4cm,angle=0,bb=0 0 270 432,keepaspectratio=true]{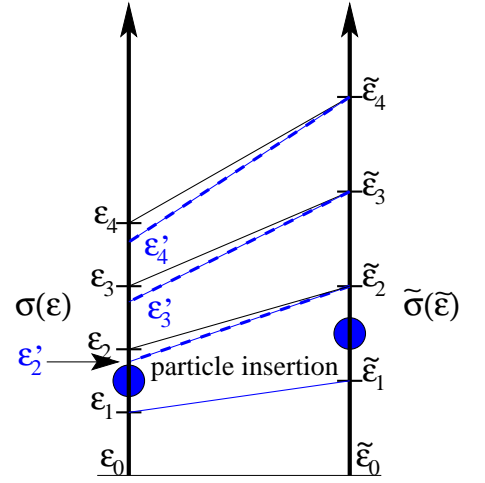}
  % FES_principle.eps: 0x0 pixel, 300dpi, 0.00x0.00 cm, bb=0 0 270 432
  \caption{(Color online) A representation of the principle of FES by using four particle species. We keep the $\tilde\epsilon$ axis fixed and the species along the $\epsilon$ axis change at the insertion of particles into the species $[\epsilon_1,\epsilon_2) \equiv [\tilde\epsilon_1,\tilde\epsilon_2)$; $\epsilon_i$ ($i=0,1,2,3,4$) correspond to $\tilde\epsilon_i$ before the insertion, whereas $\epsilon'_i$ correspond to $\tilde\epsilon_i$ after the insertion.}
  \label{FES_principle}
\end{figure}

In Eqs. (\ref{epstilgen}) and (\ref{epstilgenint}) $\tilde\epsilon$ depends on the populations $n_j$ and $n(\tilde{\epsilon'})$, respectively, so apparently they do not represent the energies of ideal particles. To view them ideal particles, we must \textit{change our perspective}. We view $\tilde\epsilon$ as the physical energy and $\epsilon$ as a function of the quasiparticle energy and the populations. This is illustrated in Fig. \ref{FES_principle}, where a particle is introduced between $\tilde\epsilon_1$ and $\tilde\epsilon_2$, or, equivalently, between $\epsilon_1$ and $\epsilon_2$. This changes the energies $\epsilon_2$, $\epsilon_3$ and $\epsilon_4$ into $\epsilon_2'$, $\epsilon_3'$ and $\epsilon_4'$, respectively, while leaving all the quasiparticle energies unchanged.

To calculate the partition function of the system I coarse-grain the $\tilde\epsilon$ axis into the intervals $[\tilde\epsilon_i,\tilde\epsilon_{i+1})$, $i=0,1,\ldots$, each interval containing 
%
%\begin{equation}
  $G(\tilde\epsilon_i,\tilde\epsilon_{i+1})\equiv\int_{\tilde\epsilon_{i}}^{\tilde\epsilon_{i+1}} \tilde\sigma(\tilde\epsilon)\,d\tilde\epsilon \label{Gij_def}$ %= \int_{\epsilon(\tilde\epsilon_1)}^{\epsilon(\tilde\epsilon_2)} \sigma(\epsilon')n(\epsilon')\,d\epsilon',
%\end{equation}
%
states and 
%
%\begin{equation}
  $N(\tilde\epsilon_{i},\tilde\epsilon_{i+1}) \equiv \int_{\tilde\epsilon_{i}}^{\tilde\epsilon_{i+1}}  \tilde\sigma(\tilde\epsilon) n(\tilde\epsilon) \,d\tilde\epsilon \label{Nij_def}$ %= \int_{\epsilon(\tilde\epsilon_1)}^{\epsilon(\tilde\epsilon_2)} \sigma(\epsilon')n(\epsilon')\,d\epsilon',
%\end{equation}
%
particles (see Fig. \ref{FES_principle}). These intervals are our species of particles.
At the insertion of particles into one of the species, say at energy $\epsilon_I$, all the species along the $\epsilon$ axis, above $\epsilon_I$, change, producing a change of the numbers of states, $G_i$, as represented in Fig. \ref{FES_principle}. This change of $G_i$ gives rise to FES \cite{PhysLettA.372.5745.2008.Anghel,PhysLettA.376.892.2012.Anghel,RJP.54.281.2009.Anghel}. 

The FES parameters were calculated in \cite{PhysLettA.372.5745.2008.Anghel,PhysLettA.376.892.2012.Anghel} and are 
\begin{equation}
\alpha_{\tilde\epsilon\tilde\epsilon_I} = \delta\epsilon\frac{d}{d\epsilon}\left\{\frac{\sigma(\epsilon)[V(\epsilon,\epsilon_I)+ f(\tilde\epsilon,\tilde\epsilon_I)]}{1+\int_{0}^{\epsilon} \frac{\partial V(\epsilon,\epsilon')}{\partial\epsilon}\sigma(\epsilon')n(\epsilon')\,d\epsilon'} \right\} \label{alpha_gen}
\end{equation}
where $\epsilon>\epsilon_I$ and $\delta\epsilon$ is the dimension of the species along the $\epsilon$ axis (e.g. $\delta\epsilon=\epsilon_{i+1}-\epsilon_i$). The function $f(\tilde\epsilon,\tilde\epsilon_I)$ is determined by the variation of $\epsilon$ at the insertion of a particle at energy $\epsilon_I<\epsilon$, written as the functional derivative, $\delta\epsilon/\delta\rho(\epsilon_I)$, 
\begin{equation}
f(\tilde\epsilon,\tilde\epsilon_I) = \int_{\epsilon_i}^{\epsilon}\frac{\partial V(\epsilon,\epsilon')}{\partial\epsilon'}\sigma(\epsilon')n(\epsilon') \left[\frac{\delta \epsilon'}{\delta\rho(\epsilon_I)}\right]_{\{\rho(\tilde\epsilon)\}}\,d\epsilon' , \label{fepsepsp_def}
\end{equation}
$\rho(\epsilon_I)$ being the density of particles along the $\epsilon$ axis, at $\epsilon_I$.

If $\epsilon=\epsilon_I$, i.e. the particle is inserted into the species where the FES parameter is calculated, then 
\begin{equation}
\tilde\alpha_{\tilde\epsilon_I,\epsilon_I} =  1+ \frac{V(\epsilon_I,\epsilon_I) \sigma(\epsilon_I)}{1+ \int_0^{\epsilon_I} \frac{\partial V(\epsilon_I,\epsilon')}{\partial\epsilon_I}\sigma(\epsilon')n(\epsilon')\,d\epsilon'} . \label{alpha_eps_eps}
\end{equation}

Once the FES parameters are known, all the thermodynamics follows by the established procedure \cite{PhysRevLett.73.922.1994.Wu,PhysRevLett.73.2150.1994.Isakov,JPhysA.40.F1013.2007.Anghel,EPL.87.60009.2009.Anghel,PhysRevLett.104.198901.2010.Anghel,PhysRevLett.104.198902.2010.Wu}.

Therefore the gas of quasiparticles, with the quasiparticle energies given by (\ref{epstilgen}) and the FES parameters (\ref{alpha_gen}) is an ideal gas, of total energy equal to the energy of the original interacting particle system and therefore with identical thermodynamic properties. 

\section{Conclusions}

I analyzed the thermodynamic properties of an interacting particle system described in the Fermi liquid theory (FLT). The Landau's quasiparticles may provide a good description of the thermodynamic properties of the system only in the mean field approximation, i.e. when the interaction between the particles contribute to the total energy of the system by a quantity that depends only on the total particle number and not on the temperature. In such a situation the thermodynamics of the system is well described by the ideal particle part of the Hamiltonian. 

If the interaction between particles contribute with a temperature dependend term to the total energy of the system, then the gas of Landau's quasiparticles do not provide an accurate description of the thermal properties of the system. For this situation I introduce another set of quasiparticles. In the new description the total energy of the system is identical to the energy of the quasiparticle system and therefore the thermodynamic quantities of the two systems are also identical. 

The new quasiparticles are ideal and obey fractional exclusion statistics. 

I consider that the fractional exclusion statistics is the paradigm in which systems of interacting particles can be described as ideal gases. 

\section*{Acknowledgements}

Discussions with dr. Gerhard M\"uller and dr. Alexandru Nemne\c s are gratefully acknowledged. The work was supported by the Romanian National Authority for Scientific Research CNCS-UEFISCDI projects PN-II-ID-PCE-2011-3-0960 and PN09370102/2009. The travel support from the Romania-JINR Dubna collaboration project Titeica-Markov and project N4063 are gratefully acknowledged.

%\bibliography{/home/dragos/general}

\begin{thebibliography}{29}
\expandafter\ifx\csname natexlab\endcsname\relax\def\natexlab#1{#1}\fi
\expandafter\ifx\csname bibnamefont\endcsname\relax
  \def\bibnamefont#1{#1}\fi
\expandafter\ifx\csname bibfnamefont\endcsname\relax
  \def\bibfnamefont#1{#1}\fi
\expandafter\ifx\csname citenamefont\endcsname\relax
  \def\citenamefont#1{#1}\fi
\expandafter\ifx\csname url\endcsname\relax
  \def\url#1{\texttt{#1}}\fi
\expandafter\ifx\csname urlprefix\endcsname\relax\def\urlprefix{URL }\fi
\providecommand{\bibinfo}[2]{#2}
\providecommand{\eprint}[2][]{\url{#2}}

\bibitem[{\citenamefont{Pines and Nozi{\'e}res}(1966)}]{PinesNozieres:book}
\bibinfo{author}{\bibfnamefont{D.}~\bibnamefont{Pines}} \bibnamefont{and}
  \bibinfo{author}{\bibfnamefont{P.}~\bibnamefont{Nozi{\'e}res}},
  \emph{\bibinfo{title}{The Theory of Quantum Liquids}} (\bibinfo{publisher}{W.
  A. Benjamin, Inc., New York, Amsterdam}, \bibinfo{year}{1966}).

\bibitem[{\citenamefont{Haldane}(1991)}]{PhysRevLett.67.937.1991.Haldane}
\bibinfo{author}{\bibfnamefont{F.~D.~M.} \bibnamefont{Haldane}},
  \bibinfo{journal}{Phys. Rev. Lett.} \textbf{\bibinfo{volume}{67}},
  \bibinfo{pages}{937} (\bibinfo{year}{1991}).

\bibitem[{\citenamefont{Murthy and
  Shankar}(1994)}]{PhysRevLett.73.3331.1994.Murthy}
\bibinfo{author}{\bibfnamefont{M.~V.~N.} \bibnamefont{Murthy}}
  \bibnamefont{and} \bibinfo{author}{\bibfnamefont{R.}~\bibnamefont{Shankar}},
  \bibinfo{journal}{Phys. Rev. Lett.} \textbf{\bibinfo{volume}{73}},
  \bibinfo{pages}{3331} (\bibinfo{year}{1994}).

\bibitem[{\citenamefont{Hansson et~al.}(1996)\citenamefont{Hansson, Leinaas,
  and Viefers}}]{NuclPhysB470.291.1996.Hansson}
\bibinfo{author}{\bibfnamefont{T.}~\bibnamefont{Hansson}},
  \bibinfo{author}{\bibfnamefont{J.}~\bibnamefont{Leinaas}}, \bibnamefont{and}
  \bibinfo{author}{\bibfnamefont{S.}~\bibnamefont{Viefers}},
  \bibinfo{journal}{Nucl. Phys. B} \textbf{\bibinfo{volume}{470}},
  \bibinfo{pages}{291} (\bibinfo{year}{1996}).

\bibitem[{\citenamefont{Isakov and
  Viefers}(1997)}]{IntJModPhysA12.1895.1997.Isakov}
\bibinfo{author}{\bibfnamefont{S.}~\bibnamefont{Isakov}} \bibnamefont{and}
  \bibinfo{author}{\bibfnamefont{S.}~\bibnamefont{Viefers}},
  \bibinfo{journal}{Int. J. Mod. Phys. A} \textbf{\bibinfo{volume}{12}},
  \bibinfo{pages}{1895} (\bibinfo{year}{1997}).

\bibitem[{\citenamefont{Bhaduri et~al.}(2000)\citenamefont{Bhaduri, Reimann,
  Viefers, Choudhury, and Srivastava}}]{JPhysB33.3895.2000.Bhaduri}
\bibinfo{author}{\bibfnamefont{R.~K.} \bibnamefont{Bhaduri}},
  \bibinfo{author}{\bibfnamefont{S.~M.} \bibnamefont{Reimann}},
  \bibinfo{author}{\bibfnamefont{S.}~\bibnamefont{Viefers}},
  \bibinfo{author}{\bibfnamefont{A.~G.} \bibnamefont{Choudhury}},
  \bibnamefont{and} \bibinfo{author}{\bibfnamefont{M.~K.}
  \bibnamefont{Srivastava}}, \bibinfo{journal}{J. Phys. B}
  \textbf{\bibinfo{volume}{33}}, \bibinfo{pages}{3895} (\bibinfo{year}{2000}).

\bibitem[{\citenamefont{Hansson et~al.}(2001)\citenamefont{Hansson, Leinaas,
  and Viefers}}]{PhysRevLett.86.2930.2001.Hansson}
\bibinfo{author}{\bibfnamefont{T.~H.} \bibnamefont{Hansson}},
  \bibinfo{author}{\bibfnamefont{J.~M.} \bibnamefont{Leinaas}},
  \bibnamefont{and} \bibinfo{author}{\bibfnamefont{S.}~\bibnamefont{Viefers}},
  \bibinfo{journal}{Phys. Rev. Lett.} \textbf{\bibinfo{volume}{86}},
  \bibinfo{pages}{2930} (\bibinfo{year}{2001}).

\bibitem[{\citenamefont{Anghel}(2009{\natexlab{a}})}]{RJP.54.281.2009.Anghel}
\bibinfo{author}{\bibfnamefont{D.~V.} \bibnamefont{Anghel}},
  \bibinfo{journal}{Rom. J. Phys.} \textbf{\bibinfo{volume}{54}},
  \bibinfo{pages}{281} (\bibinfo{year}{2009}{\natexlab{a}}),
  \bibinfo{note}{arXiv:0804.1474}.

\bibitem[{\citenamefont{Isakov}(1994)}]{PhysRevLett.73.2150.1994.Isakov}
\bibinfo{author}{\bibfnamefont{S.~B.} \bibnamefont{Isakov}},
  \bibinfo{journal}{Phys. Rev. Lett.} \textbf{\bibinfo{volume}{73}},
  \bibinfo{pages}{2150} (\bibinfo{year}{1994}).

\bibitem[{\citenamefont{Bernard and Wu}(1995)}]{NewDevIntSys.1995.Bernard}
\bibinfo{author}{\bibfnamefont{D.}~\bibnamefont{Bernard}} \bibnamefont{and}
  \bibinfo{author}{\bibfnamefont{Y.~S.} \bibnamefont{Wu}}, in
  \emph{\bibinfo{booktitle}{New Developments on Integrable Systems and
  Long-Ranged Interaction Models}}, edited by
  \bibinfo{editor}{\bibfnamefont{M.~L.} \bibnamefont{Ge}} \bibnamefont{and}
  \bibinfo{editor}{\bibfnamefont{Y.~S.} \bibnamefont{Wu}}
  (\bibinfo{publisher}{World Scientific, Singapore}, \bibinfo{year}{1995}),
  p.~\bibinfo{pages}{10}, \bibinfo{note}{cond-mat/9404025}.

\bibitem[{\citenamefont{Sutherland}(1997)}]{PhysRevB.56.4422.1997.Sutherland}
\bibinfo{author}{\bibfnamefont{B.}~\bibnamefont{Sutherland}},
  \bibinfo{journal}{Phys. Rev. B} \textbf{\bibinfo{volume}{56}},
  \bibinfo{pages}{4422} (\bibinfo{year}{1997}).

\bibitem[{\citenamefont{Iguchi and
  Sutherland}(2000)}]{PhysRevLett.85.2781.2000.Iguchi}
\bibinfo{author}{\bibfnamefont{K.}~\bibnamefont{Iguchi}} \bibnamefont{and}
  \bibinfo{author}{\bibfnamefont{B.}~\bibnamefont{Sutherland}},
  \bibinfo{journal}{Phys. Rev. Lett.} \textbf{\bibinfo{volume}{85}},
  \bibinfo{pages}{2781} (\bibinfo{year}{2000}).

\bibitem[{\citenamefont{Potter et~al.}(2007{\natexlab{a}})\citenamefont{Potter,
  M{\"u}ller, and Karbach}}]{PhysRevE.75.61120.2007.Potter}
\bibinfo{author}{\bibfnamefont{G.~G.} \bibnamefont{Potter}},
  \bibinfo{author}{\bibfnamefont{G.}~\bibnamefont{M{\"u}ller}},
  \bibnamefont{and} \bibinfo{author}{\bibfnamefont{M.}~\bibnamefont{Karbach}},
  \bibinfo{journal}{Phys. Rev. E} \textbf{\bibinfo{volume}{75}},
  \bibinfo{pages}{61120} (\bibinfo{year}{2007}{\natexlab{a}}).

\bibitem[{\citenamefont{Potter et~al.}(2007{\natexlab{b}})\citenamefont{Potter,
  M{\"u}ller, and Karbach}}]{PhysRevE.76.61112.2007.Potter}
\bibinfo{author}{\bibfnamefont{G.~G.} \bibnamefont{Potter}},
  \bibinfo{author}{\bibfnamefont{G.}~\bibnamefont{M{\"u}ller}},
  \bibnamefont{and} \bibinfo{author}{\bibfnamefont{M.}~\bibnamefont{Karbach}},
  \bibinfo{journal}{Phys. Rev. E} \textbf{\bibinfo{volume}{76}},
  \bibinfo{pages}{61112} (\bibinfo{year}{2007}{\natexlab{b}}).

\bibitem[{\citenamefont{Anghel}(2010{\natexlab{a}})}]{EPL.90.10006.2010.Anghel}
\bibinfo{author}{\bibfnamefont{D.~V.} \bibnamefont{Anghel}},
  \bibinfo{journal}{EPL} \textbf{\bibinfo{volume}{90}}, \bibinfo{pages}{10006}
  (\bibinfo{year}{2010}{\natexlab{a}}), \bibinfo{note}{arXiv:0909.0030}.

\bibitem[{\citenamefont{Liu et~al.}(2011)\citenamefont{Liu, Lu, M\"{u}ller, and
  Karbach}}]{PhysRevE.84.021136.2011.Liu}
\bibinfo{author}{\bibfnamefont{D.}~\bibnamefont{Liu}},
  \bibinfo{author}{\bibfnamefont{P.}~\bibnamefont{Lu}},
  \bibinfo{author}{\bibfnamefont{G.}~\bibnamefont{M\"{u}ller}},
  \bibnamefont{and} \bibinfo{author}{\bibfnamefont{M.}~\bibnamefont{Karbach}},
  \bibinfo{journal}{Phys. Rev. E} \textbf{\bibinfo{volume}{84}},
  \bibinfo{pages}{021136} (\bibinfo{year}{2011}).

\bibitem[{\citenamefont{Liu et~al.}(2012)\citenamefont{Liu, Vanasse,
  M\"{u}ller, and Karbach}}]{PhysRevE.85.011144.2012.Liu}
\bibinfo{author}{\bibfnamefont{D.}~\bibnamefont{Liu}},
  \bibinfo{author}{\bibfnamefont{J.}~\bibnamefont{Vanasse}},
  \bibinfo{author}{\bibfnamefont{G.}~\bibnamefont{M\"{u}ller}},
  \bibnamefont{and} \bibinfo{author}{\bibfnamefont{M.}~\bibnamefont{Karbach}},
  \bibinfo{journal}{Phys. Rev. E} \textbf{\bibinfo{volume}{85}},
  \bibinfo{pages}{011144} (\bibinfo{year}{2012}).

\bibitem[{\citenamefont{Anghel}(2002)}]{JPA35.7255.2002.Anghel}
\bibinfo{author}{\bibfnamefont{D.~V.} \bibnamefont{Anghel}},
  \bibinfo{journal}{J. Phys. A: Math. Gen.} \textbf{\bibinfo{volume}{35}},
  \bibinfo{pages}{7255} (\bibinfo{year}{2002}).

\bibitem[{\citenamefont{Anghel}(2008)}]{PhysLettA.372.5745.2008.Anghel}
\bibinfo{author}{\bibfnamefont{D.~V.} \bibnamefont{Anghel}},
  \bibinfo{journal}{Phys. Lett. A} \textbf{\bibinfo{volume}{372}},
  \bibinfo{pages}{5745} (\bibinfo{year}{2008}),
  \bibinfo{note}{arXiv:0710.0728}.

\bibitem[{\citenamefont{Anghel}(2012)}]{PhysLettA.376.892.2012.Anghel}
\bibinfo{author}{\bibfnamefont{D.~V.} \bibnamefont{Anghel}},
  \bibinfo{journal}{Phys. Lett. A} \textbf{\bibinfo{volume}{376}},
  \bibinfo{pages}{892} (\bibinfo{year}{2012}).

\bibitem[{\citenamefont{Lewin}(1958)}]{Lewin:book}
\bibinfo{author}{\bibfnamefont{L.}~\bibnamefont{Lewin}},
  \emph{\bibinfo{title}{Dilogarithms and Associated Functions}}
  (\bibinfo{publisher}{Macdonald, London}, \bibinfo{year}{1958}).

\bibitem[{\citenamefont{Lee}(2009)}]{ActaPhysicaPolonicaB.40.1279.2009.Lee}
\bibinfo{author}{\bibfnamefont{M.~H.} \bibnamefont{Lee}},
  \bibinfo{journal}{Acta Physica Polonica B} \textbf{\bibinfo{volume}{40}},
  \bibinfo{pages}{1279} (\bibinfo{year}{2009}).

\bibitem[{\citenamefont{Stone}(1994)}]{Stone:book}
\bibinfo{author}{\bibfnamefont{M.}~\bibnamefont{Stone}},
  \emph{\bibinfo{title}{Bosonisation}} (\bibinfo{publisher}{World Scientific},
  \bibinfo{year}{1994}).

\bibitem[{\citenamefont{Pethic and Carneiro}(1973)}]{PRA7.304.1973.Pethic}
\bibinfo{author}{\bibfnamefont{J.~C.} \bibnamefont{Pethic}} \bibnamefont{and}
  \bibinfo{author}{\bibfnamefont{G.~M.} \bibnamefont{Carneiro}},
  \bibinfo{journal}{Phys. Rev. A} \textbf{\bibinfo{volume}{7}},
  \bibinfo{pages}{304} (\bibinfo{year}{1973}).

\bibitem[{\citenamefont{Wu}(1994)}]{PhysRevLett.73.922.1994.Wu}
\bibinfo{author}{\bibfnamefont{Y.-S.} \bibnamefont{Wu}},
  \bibinfo{journal}{Phys. Rev. Lett.} \textbf{\bibinfo{volume}{73}},
  \bibinfo{pages}{922} (\bibinfo{year}{1994}).

\bibitem[{\citenamefont{Anghel}(2007)}]{JPhysA.40.F1013.2007.Anghel}
\bibinfo{author}{\bibfnamefont{D.~V.} \bibnamefont{Anghel}},
  \bibinfo{journal}{J. Phys. A: Math. Theor.} \textbf{\bibinfo{volume}{40}},
  \bibinfo{pages}{F1013} (\bibinfo{year}{2007}),
  \bibinfo{note}{arXiv:0710.0724}.

\bibitem[{\citenamefont{Anghel}(2009{\natexlab{b}})}]{EPL.87.60009.2009.Anghel}
\bibinfo{author}{\bibfnamefont{D.~V.} \bibnamefont{Anghel}},
  \bibinfo{journal}{EPL} \textbf{\bibinfo{volume}{87}}, \bibinfo{pages}{60009}
  (\bibinfo{year}{2009}{\natexlab{b}}), \bibinfo{note}{arXiv:0906.4836}.

\bibitem[{\citenamefont{Anghel}(2010{\natexlab{b}})}]{PhysRevLett.104.198901.2%
010.Anghel}
\bibinfo{author}{\bibfnamefont{D.~V.} \bibnamefont{Anghel}},
  \bibinfo{journal}{Phys. Rev. Lett.} \textbf{\bibinfo{volume}{104}},
  \bibinfo{pages}{198901} (\bibinfo{year}{2010}{\natexlab{b}}).

\bibitem[{\citenamefont{Wu}(2010)}]{PhysRevLett.104.198902.2010.Wu}
\bibinfo{author}{\bibfnamefont{Y.-S.} \bibnamefont{Wu}},
  \bibinfo{journal}{Phys. Rev. Lett.} \textbf{\bibinfo{volume}{104}},
  \bibinfo{pages}{198902} (\bibinfo{year}{2010}).

\end{thebibliography}
%\bibliographystyle{unsrt}

\end{document}